\documentclass[aps, prl, 10pt, twocolumn, 
superscriptaddress, amsmath, amssymb, letterpaper]{revtex4-1}
\usepackage[pdftex]{graphicx}
\usepackage[ngerman,english]{babel}
\usepackage[ansinew]{inputenc}
\usepackage{natbib}
\usepackage{color}
\usepackage{siunitx}
\usepackage{calc}
\usepackage[colorlinks=true, pdfstartview=FitV, linkcolor=blue,
citecolor=blue, urlcolor=blue]{hyperref}

\newcommand{\ie}{{i.e. }}

\definecolor{green2}{rgb}{.0, .58, 0}

\begin{document}

\title{Magnetic force sensing using a self-assembled nanowire}

\author{N.~Rossi} \affiliation{Department of Physics, University of Basel, 4056 Basel, Switzerland}

\author{B.~Gross} \affiliation{Department of Physics, University of Basel, 4056 Basel, Switzerland}

\author{F.~Dirnberger} \affiliation{Institut f\"{u}r Experimentelle und Angewandte Physik, Universit\"{a}t Regensburg, D-93040 Regensburg, Germany}

\author{D.~Bougeard} \affiliation{Institut f\"{u}r Experimentelle und Angewandte Physik, Universit\"{a}t Regensburg, D-93040 Regensburg, Germany}

\author{M.~Poggio} \affiliation{Department of Physics, University of Basel, 4056 Basel, Switzerland} \email{martino.poggio@unibas.ch}

\begin{abstract} 

  We present a scanning magnetic force sensor based on an individual
  magnet-tipped GaAs nanowire (NW) grown by molecular beam epitaxy.
  Its magnetic tip consists of a final segment of single-crystal MnAs
  formed by sequential crystallization of the liquid Ga catalyst
  droplet.  We characterize the mechanical and magnetic properties of
  such NWs by measuring their flexural mechanical response in an
  applied magnetic field.  Comparison with numerical simulations
  allows the identification of their equilibrium magnetization
  configurations, which in some cases include magnetic vortices.  To
  determine a NW's performance as a magnetic scanning probe, we
  measure its response to the field profile of a lithographically
  patterned current-carrying wire.  The NWs' tiny tips and their high
  force sensitivity make them promising for imaging weak magnetic
  field patterns on the nanometer-scale, as required for mapping
  mesoscopic transport and spin textures or in nanometer-scale
  magnetic resonance.

\end{abstract} 

\maketitle



A key component in any force microscopy is the force sensor. This
device consists of a mechanical transducer, used to convert force into
displacement, and an optical or electrical displacement detector. In
magnetic force microscopy (MFM), mass-produced `top-down' Si
cantilevers with sharp tips coated by a magnetic material have been
the standard for years. Under ideal conditions, state-of-the-art MFM
can reach spatial resolutions down to 10 nm
\cite{schmid_exchange_2010}, though more typically around
$100$~nm. These conventional cantilevers are well-suited for the
measurement of the large forces and force gradients produced by
strongly magnetized samples.

The advent of nanostructures such as nanowires (NWs) and carbon
nanotubes (CNTs) grown by `bottom-up' techniques has given researchers
access to much smaller mechanical force transducers than ever before.
This reduction in size implies both a better force
sensitivity~\cite{poggio_sensing_2013} and -- potentially -- a finer
spatial resolution \cite{lisunova_optimal_2013}.  Sensitivity to small
forces provides the ability to detect weak magnetic fields and
therefore to image subtle magnetic patterns; tiny concentrated
magnetic tips have the potential to achieve nanometer-scale spatial
resolution, while also reducing the invasiveness of the tip on the
sample under investigation.  Such improvements are crucial for imaging
nanometer-scale magnetization textures such as domain walls, votices
and
skyrmions~\cite{rondin_stray-field_2013,tetienne_nanoscale_2014,tetienne_nature_2015,dovzhenko_imaging_2016};
superconducuting
vortices~\cite{thiel_quantitative_2016,pelliccione_scanned_2016};
mesoscopic transport in two-dimensional
systems~\cite{tetienne_quantum_2017}; and small ensembles of nuclear
spins~\cite{rugar_proton_2015,haberle_nanoscale_2015,degen_nanoscale_2009,poggio_force-detected_2010}.

Recent experiments have demonstrated the use of single NWs and CNTs as
sensitive scanning force
sensors~\cite{gloppe_bidimensional_2014,rossi_vectorial_2017,de_lepinay_universal_2017,siria_electron_2017}.
When clamped on one end and arranged in the pendulum geometry, i.e.\
with their long axes perpendicular to the sample surface to prevent
snapping into contact, they probe both the size and direction of weak
tip-sample forces.  NWs have been demonstrated to maintain excellent
force sensitivities around 1 $\text{aN}/\sqrt{\text{Hz}}$ near sample
surfaces ($<100$ nm), due to extremely low non-contact
friction~\cite{nichol_nanomechanical_2012}.  As a result, NW sensors
have been used as transducers in force-detected nanometer-scale
magnetic resonance imaging~\cite{nichol_nanoscale_2013} and in the
measurement of optical and electrical
forces~\cite{gloppe_bidimensional_2014,rossi_vectorial_2017,de_lepinay_universal_2017}.
Nevertheless, the integration of a magnetic tip onto a NW transducer
-- and therefore the demonstration of NW MFM -- has presented a
significant practical challenge.

Here, we demonstrate such MFM transducers using individual GaAs NWs
with integrated single-crystal MnAs tips, grown by molecular beam
epitaxy (MBE).  By monitoring each NW's flexural motion in an applied
magnetic field, we determine its mechanical and magnetic properties.
We determine the equilibrium magnetization configurations of each tip
by comparing its magnetic response with micromagnetic simulations.  In
order to determine the sensitivity and resolution of the NWs as MFM
transducers, we use them as scanning probes in the pendulum geometry.
By analyzing their response to the magnetic field produced by a
lithographically patterned current-carrying wire, we find that the
MnAs tips can be approximated as nearly perfect magnetic dipoles.  The
thermally-limited sensitivity of a typical NW to magnetic field
gradients is found to be 11 $\text{mT}/(\text{m}\sqrt{\text{Hz})}$,
which corresponds to the gradient produced by
$61\;\text{nA}/\sqrt{\text{Hz}}$ through the wire at a tip-sample
spacing of $250$~nm.


The GaAs NWs are grown on a Si(111) substrate by MBE using a
self-catalyzed Ga-assisted growth
method~\cite{colombo_ga-assisted_2008}.  A substrate temperature of
$600$~$^\circ$C allows the growth of high quality crystalline NWs,
which are typically $18 \pm 1$ $\mu$m long with a hexagonal
cross-section of $225 \pm 15$ nm in diameter.  In order to terminate
the growth with a magnetic tip, the liquid Ga catalyst droplet at the
top of the NW is heavily alloyed by a Mn flux. Then, to initiate its
crystallization, it is exposed to an As background pressure for $30$
minutes.  Under such conditions, the droplet undergoes a sequential
precipitation: first, the Ga is preferentially consumed to build pure
GaAs; next, the remaining Mn crystallizes in the form of MnAs.  It has
been shown by high-resolution transmission electron microscopy that
this growth process leads to the formation of a well-defined hexagonal
$\alpha$-MnAs wurzite crystal at the tip of a predominantly wurzite
GaAs NW, with an epitaxial relationship $[0001]\text{MnAs}\parallel
[0001]\text{GaAs}$ along the NW-axis \cite{hubmann_epitaxial_2016}.
As reported for bulk MnAs, the tip is in a hexagonal ferromagnetic
$\alpha$-phase up to the Curie temperature of about $313$ K, above
which it undergoes a structural phase transition into an orthorhombic
paramagnetic $\beta$-phase \cite{bean_magnetic_1962}.  MnAs crystals
are characterized by a strong magnetocrystalline anisotropy with $K =
-1\times 10^6$ J m$^{-3}$ along the $c$-axis (hard axis)
\cite{de_blois_magnetic_1963}.  As a result, the magnetization of the
tip will tend to lie in the plane (easy plane) orthogonal to the
$c$-axis, which -- in general -- is coincident with the NW growth
direction $\mathbf{\hat{n}}$.

\begin{figure}
  \includegraphics[]{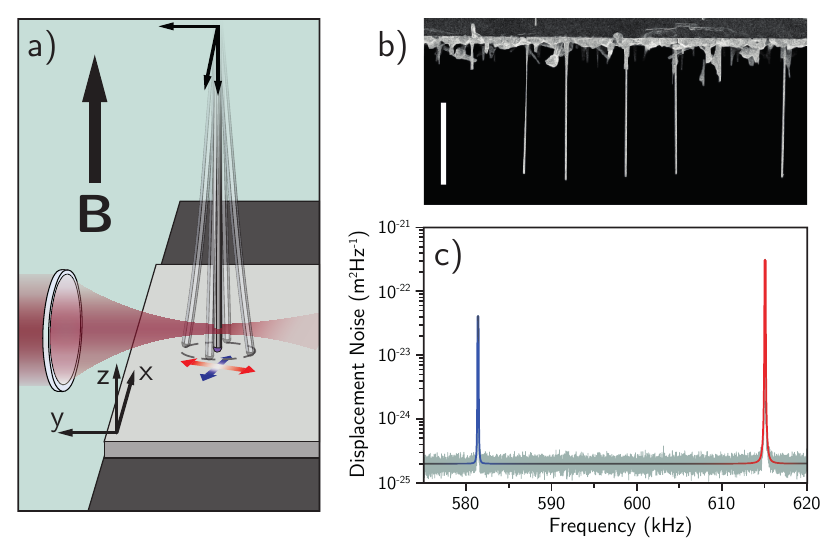}
\caption{(a) Schematic drawing of the measurement setup.  The NW is be
  positioned in the waist of the focused laser beam by means of
  piezoelectric positioners.  A second stack of positioners is used to
  approach and scan the sample of interest with respect to the NW.
  (b) Scanning electron microscope (SEM) micrograph of the measured
  chip cleaved from the growth wafer with a $10\;\mu$m scale-bar.  (c)
  Displacement noise PSD of the first order flexural modes of NW$2$,
  at $B=0$~T.  Measured data (gray) and fit (colored line)
  representing the sum of two Lorentzians and white noise background
  $S_N = 2\times10^{-25}$ m$^2$/Hz.  The fit yields the orientation of
  each mode with respect to $\mathbf{\hat{y}}$: $\phi_1 =
  +70^{\circ}$, $\phi_2= - 20^{\circ}$ and the modes' effective mass
  as 588 fg, which is close to $790$ fg expected for a GaAs cylinder
  of density $5.32$ g/cm$^3$ with hexagonal cross-section of maximal
  diameter $234$ nm and length 16.8~$\mu$m.  The extracted resonance
  frequency, spring constant, and intrinsic quality factor of the
  first (second) mode are $f_{0_1} = 581$ kHz, $k_1 = 7.8$ mN/m and
  $Q_1 = 44650$ ($f_{0_2} = 615$ kHz, $k_2 = 8.8$ mN/m and $Q_2 =
  48456$).  }
\label{fig:1}
\end{figure}

The sample chip is cleaved directly from the Si wafer used for the
NWs' growth.  Using a micromanipulator under an optical microscope, we
remove excess NWs to leave a single row of isolated and vertically
standing NWs in proximity of the cleaved edge (Fig \ref{fig:1}(b)).
The chip is then loaded into a custom-built scanning probe microscope,
which includes piezoelectric positioners to align a single NW within
the focus of a fiber-coupled optical interferometer used to detect its
mechanical motion \cite{nichol_displacement_2008}.  A second set of
piezoelectric positioners enables the approach and scanning of the NW
transducer over a sample of interest, as shown schematically in
Fig.~\ref{fig:1}(a).  The microscope is enclosed in a high-vacuum
chamber at a pressure of $10^{ -7}$ mbar and inserted in the bore of a
superconducting magnet at the bottom of a liquid $^4$He bath cryostat.
All the data presented here have been measured at a temperature $T
=4.2$ K.  In order to characterize the magnetic properties of the NWs,
we apply a magnetic field $\mathbf{B}$ up to $\pm8$ T approximately
parallel to $\mathbf{\hat{n}}$.

%

For the purposes of this work, we restrict our attention to the two
fundamental flexural eigenmodes of the NWs, which oscillate along
orthogonal directions and are shown schematically in
Fig.~\ref{fig:1}(a).  The coupling between the NW and the thermal bath
results in a Langevin force equally driving both mechanical modes.
Fig.~\ref{fig:1}(c) shows a calibrated power spectral density (PSD) of
the NW displacement noise, where the two resonance peaks correspond to
the two orthogonally polarized modes. Such a measurement shows the
projection of the modes' 2D thermal motion along the interferometric
measurement axis $\mathbf{\hat{m}}$, determined by the position of the
NW in the optical waist (see Methods).  Typical resonance frequencies
range from 500 to 700~kHz with quality factors between $2\times10^4$
and $5\times10^4$.  For each NW, the doublet modes are completely
decoupled by a frequency splitting $\delta$ of several hundred times
the peak linewidth.  In fact, it has been shown that even very small
($<1\%$) cross-sectional or clamping asymmetries can split the modes
by several linewidths~\cite{cadeddu_time-resolved_2016}.
Nevertheless, the quality factors of the doublet modes differ by less
than $1\%$.  The spring constants extracted from fits to the PSD for
each flexural mode are on the order of $10\;\text{mN}/\text{m}$,
yielding mechanical dissipations and thermally-limited force
sensitivities down to $50$ pg/s and few $\text{aN}/\sqrt{\text{Hz}}$,
respectively.


\begin{figure} [t!] 
	\includegraphics[]{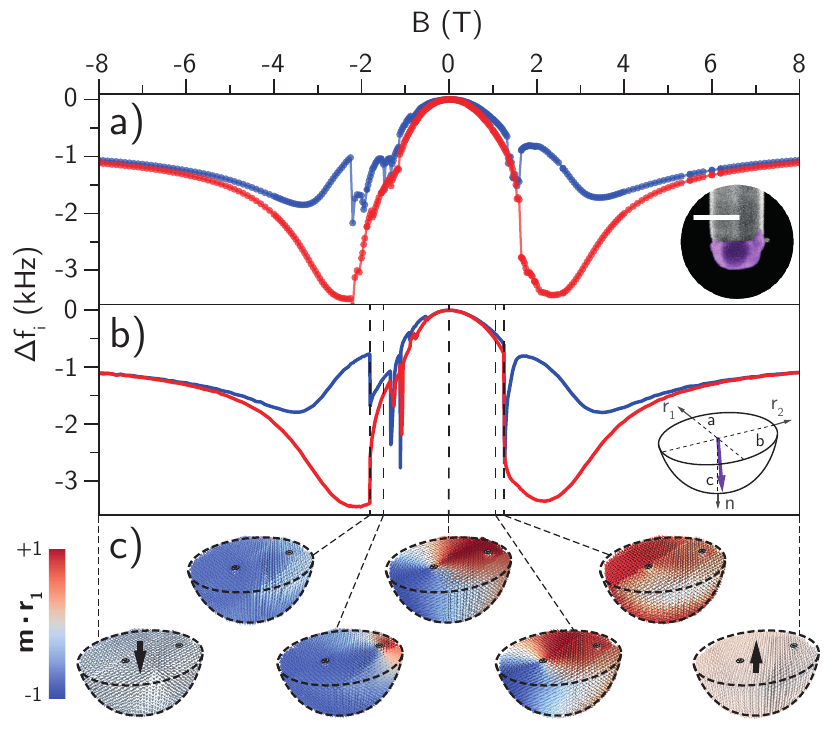}
	\caption{DCM measurement of NW$1$ revealing the existence of a
		magnetic vortex in the MnAs tip.  (a) Plot of the $\Delta f_i (B)$
		for $i = 1$ ($i = 2$) in blue (red), extracted at each $B$ from the
		thermal displacement PSD.  $f_{0_1} = 613$ kHz, $f_{0_2} = 636$ kHz
		and $k_1 = 7.7$ mN/m, $k_2 = 8.3$ mN/m.  $B$ is stepped from
		positive to negative field values.  Inset: false-color SEM close-up
		of NW$1$'s MnAs tip with a 200~nm scale-bar.  (b) Plot of the
		corresponding $\Delta f_i (B)$ simulated for a half-ellipsoid.
		Inset: schematic diagram of the half-ellipsoid geometry used to
		approximate the MnAs tip, where $a=85$ nm, $b=90$ nm, $c=95$ nm.
		The simulation uses parameters for MnAs from the literature and a
		hard axis direction (purple arrow) given by $\theta_K =
		4.2^{\circ}$ and $\phi_K = 45.5^{\circ}$.  (c) Simulated
		magnetization configurations of the MnAs tip at the values of $B$
		indicated by the dashed lines.  Each cone, associated with a
		discretized volume, is color-coded with the magnitude of the
		magnetization component along $\mathbf{\hat{r}}_1$.  Black dots and
		crosses indicate the position and direction of the sites of pinned
		magnetization.} 
	\label{fig:2}
\end{figure}

\begin{figure}[t!]
	\includegraphics[]{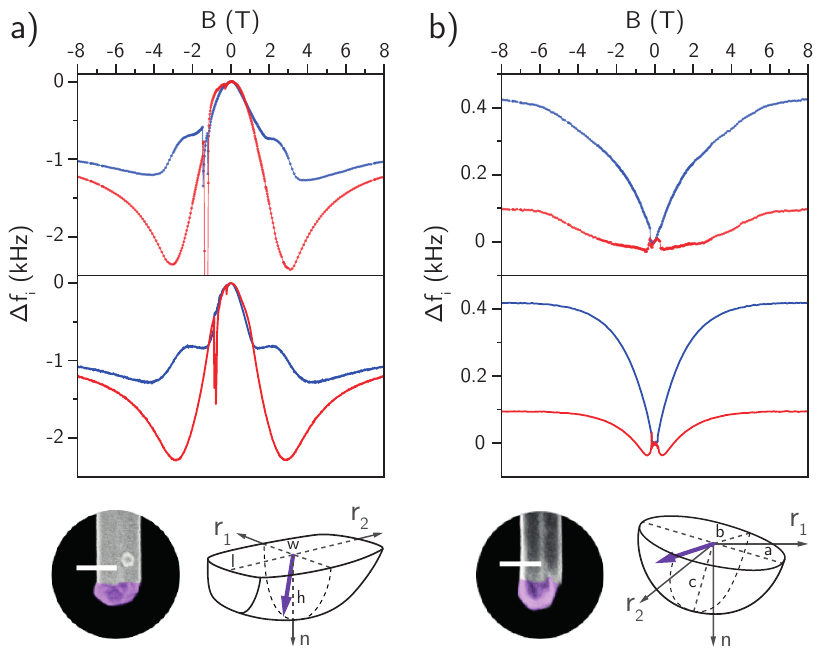} 
	\caption{DCM measurements of (a) NW$2$ and (b) NW$3$.  The upper
		plots show the measurements of $\Delta f_i (B)$ and the lower
		plots show the corresponding simulations for $i=1$ ($i=2$) in blue
		(red).  $B$ is stepped from positive to negative field values.  At
		the bottom, SEM close-ups of the NWs with MnAs tips in false color
		(scale-bar $200$ nm) along with a sketch of the tip geometry used
		in the simulation.  (a) For NW$2$, the frequency shifts have been
		calculated by using its mechanical characteristics reported in Fig
		\ref{fig:1}(c).  The direction of the hard axis (purple arrow) is
		set to $\theta_K = 8^{\circ}$ and $\phi_K = -9.5^{\circ}$.  The
		asymmetric geometry is obtained by truncating an ellipsoid, and
		resulting widths are $l = 210$~nm and $w=110$~nm while depth is
		$100$ nm.  (b) For NW$3$, resonance frequencies and spring
		constants are $f_{0_1} = 435$ kHz, $f_{0_2} = 447$ kHz and $k_1 =
		14.3$ mN/m, $k_2 = 15$ mN/m, respectively.  The hard axis (purple
		arrow) is nearly perpendicular to $\mathbf{\hat{n}}$ with
		$\theta_K = 68.3^{\circ}$ and $\phi_K = 142.2^{\circ}$.  The tip
		is approximated as an elongated ellipsoid with $a = b = 85$~nm and
		$c = 90$~nm, which is tilted respect to the modes' reference
		system by $\theta_t = 20^{\circ}$ and $\phi_t = 210^{\circ}$.}
	\label{fig:3}
\end{figure}

We exploit this high mechanical sensitivity to probe the magnetization
of each individual magnetic tip. As in dynamic cantilever magnetometry
(DCM) \cite{rossel_active_1996, harris_integrated_1999,
  stipe_magnetic_2001}, we can extract magnetic properties of each
MnAs tip from the mechanical response of the NW to a uniform external
magnetic field $\mathbf{B}$.  In such a field, the resonance frequency
of each orthogonal flexural mode $f_i$ ($i = 1, 2$) is modified by the
curvature of the system's magnetic energy $E_m$ with respect to
rotations $\theta_i$ about its oscillation axis.  The resulting
frequency shift $\Delta f_i = f_i - f_{0_i}$, where $f_{0_i}$ is the
resonance frequency at $B=0$, is given by
\begin{equation}
  \Delta f_i = \frac{f_{0_i}}{2 k_i l_e^2} \left ( \left.  \frac{\partial^2 
        E_m}{\partial \theta_i^2} \right|_{\theta_i=0} \right ),
\label{DCM_equation}
\end{equation}
where $k_i$ is the NW's spring constant and $l_e$ is its effective
length~\cite{gross_dynamic_2016,mehlin_observation_2018}.

\begin{figure*}[t!]
	\includegraphics[width=\textwidth]{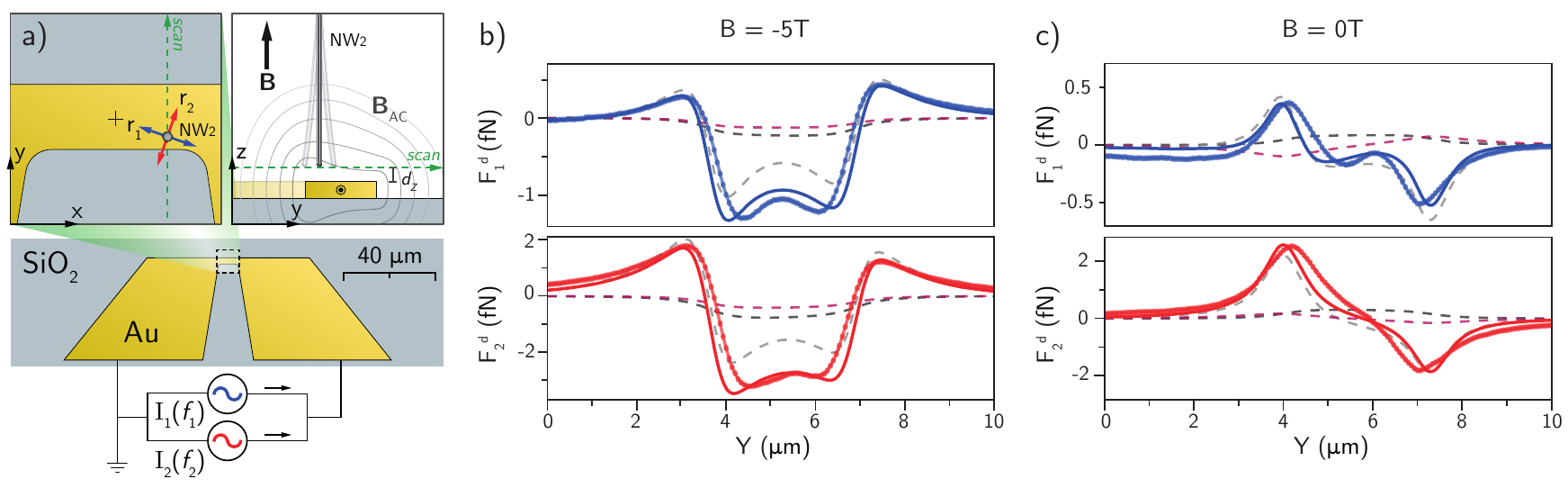}
	\caption{Characterization of NW as magnetic scanning probe.  (a)
		Schematic drawing of the bottom sample consisting of a
		3-$\mu$m-wide, 6-$\mu$m-long, and $240$-nm-thick Au wire on an
		insulating SiO$_2$ substrate.  NW$2$ is approached at a tip-sample
		spacing $d_z = 250$ nm and scanned across the constriction.  (b-c)
		Plots of the measured (dotted line) and calculated (solid line)
		forces $F_i$ driving the first (blue) and the second (red) mode
		over a $10\;\mu$m long line scan.  For each plot three distinct
		drive contributions are shown as dashed lines: the monopole
		(black), dipole (gray) and torque (magenta) terms.  The simulated
		response is fit by setting $q_0 = -1.4 \times10^{-10}$ A$\cdot$m,
		$|\mathbf{m}|=9.4\times 10^{-16}$ A$\cdot$m$^2$ with
		$\theta_{\mathbf{m}} = 2^{\circ}, \phi_{\mathbf{m}} = 20^{\circ}$
		for the high magnetic field case (b) and $q_0 = 5.5
		\times10^{-11}$ A$\cdot$m, $|\mathbf{m}|=6.8\times 10^{-16}$
		A$\cdot$m$^2$ with $\theta_{\mathbf{m}} = 87^{\circ},
		\phi_{\mathbf{m}} = 40^{\circ}$ for the zero field case (c).  }
	\label{fig:4}
\end{figure*}

We perform measurements of $\Delta f_i (B)$ on several NWs by
recording the thermal displacement PSD of their doublet modes as a
function of $B$.  For nearly all investigated NWs ($11$ out of $12$),
$\Delta f_i (B)$ is negative for all applied fields (e.g.\
Fig.~\ref{fig:2}(a) and Fig.~\ref{fig:3}(a)).  In general, negative
values of $\Delta f_i$ correspond to a local maximum in
$E_m(\theta_i)$ with respect to $\theta_i$.  This behavior is
consistent with $\mathbf{B}$ being aligned along the magnetic hard
axis of the MnAs tip, which should be along the NW growth-axis.
Fig.~\ref{fig:2}(a) shows a particularly ideal magnetic response, in
which the high-field frequency shift of both modes asymptotically
approaches the same negative value.  This behavior indicates a MnAs
particle with a hard axis along $\mathbf{\hat{n}}$ and no preferred
easy axis in the $r_1 r_2$-plane.

In order to gain a deeper understanding of the DCM signal, we carry
out simulations of the MnAs tips using
\textit{Mumax3}~\cite{vansteenkiste_design_2014}, which employs the
Landau-Lifshitz-Gilbert micromagnetic formalism using
finite-difference discretization.  For each value of $B$, the
simulations determine the equilibrium magnetization configuration of
the MnAs particle and the corresponding values of $\Delta f_i$ (see
Methods).  The geometry of the MnAs tip is estimated by SEM and set
within the simulation with respect to the two oscillation directions
of the modes $\mathbf{\hat{r}}_1 , \mathbf{\hat{r}}_2$ and the NW axis
$\mathbf{\hat{n}} = \mathbf{\hat{r}}_1 \times \mathbf{\hat{r}}_2$

The DCM response of the MnAs tip measured in Fig.~\ref{fig:2}(a) and
shown in the inset is simulated by approximating its shape as a half
ellipsoid, with dimensions given in the inset of Fig.~\ref{fig:2}(b)
and its caption.  The excellent agreement between the simulated and
measured $\Delta f_i (B)$, both plotted in Fig.~\ref{fig:2}(b), allows
us to precisely determine the direction of the magnetic hard axis.  As
expected, this axis is found to be nearly along $\mathbf{\hat{n}}$:
just $\theta_K = 2.5^{\circ}$ away from $\mathbf{\hat{n}}$ and $\phi_K
= 19.5^{\circ}$ from $\mathbf{\hat{r}_1}$.  Furthermore, as shown in
Fig.~\ref{fig:2}(c), the simulations relate a specific magnetization
configuration to each value of $B$.  In this particular case, a stable
vortex configuration in the easy plane is seen to enter (exit) from
the edge in correspondence with the abrupt discontinuities in the
eigenmodes' frequencies around $+2$ T ($-2$ T).  Between these two
fields, the vortex core moves from one side to the other, inducing
several discontinuities in $\Delta f_i(B)$.  The smoothness of the
measured frequency shifts around $B=0$~T indicates pinning of the
vortex and is well-reproduced in the simulation by the introduction of
two sites of pinned magnetization (see Supplementary Information)).

Most measured NWs present DCM curves as shown in Fig.~\ref{fig:3}(a)
($10$ out of $12$).  Despite the similarity of these curves to those
shown in Fig.~\ref{fig:2}(a), no sharp discontinuity is observed upon
sweeping $B$ down from saturation (forward applied field).
Furthermore, the high-field frequency shift of both modes does not
asymptotically approach the same negative value as in
Fig.~\ref{fig:2}(a).  Both of these effects can be explained by taking
into account magnetic shape anisotropy in the MnAs tips.  Despite the
nearly perfect symmetry of NW$1$'s tip, most of the crystallized MnAs
droplets are asymmetric in the $r_1r_2$-plane.  This asymmetry
introduces an effective magnetic easy axis in the $r_1r_2$-plane.  In
fact, the measured $\Delta f_i (B)$ shown in Fig.~\ref{fig:3}(a) are
well-reproduced by a simulation that takes into a account the geometry
of NW$2$'s MnAs tip as observed by SEM.  While small refinements in
the microscopic geometry, which often cannot be confirmed by the SEM,
affect how well the the simulation matches every detail of the
measured $\Delta f_i (b)$ (see Supplementary Information), the precise
orientation of the hard axis and the direction of the effective shape
anisotropy in the $r_1r_2$-plane sensitively determine the curves'
overall features (e.g.\ their high field asymptotes and shape).

In general, simulations show that shape anisotropy restricts the field
range for a stable magnetic vortex to reverse applied field.  In small
forward applied field and in remanence, the magnetization evolves
through a configuration with a net magnetic dipole in the
$r_1r_2$-plane. Only upon application of a reverse field, does this
configuration smoothly transform into a vortex, resulting -- for NW2
-- in a subtle dip in $\Delta f_i (B)$ around $B = -0.3$~T.  At a
reverse field close to $B = -2$~T, an abrupt jump indicates the
vortex's exit and the appearance of a single-domain state, which
eventually turns toward $\mathbf{B}$.  This analysis indicates that
NW$2$'s tip -- as well as the majority of the MnAs tips -- present a
dipole-like remanent configuration pointing in the $r_1r_2$-plane,
rather than vortex-like configuration with a core pointing along
$\mathbf{\hat{n}}$, as in NW$1$.  Such remanent magnetic dipoles have
been already observed by MFM in similar
tips~\cite{ramlan_ferromagnetic_2006,hubmann_epitaxial_2016}.

In rare cases (1 of 12), such as the one reported for NW$3$ in
Fig.~\ref{fig:3}(b), we measure mostly positive $\Delta f_i (B)$ with
different high-field asymptotes for each eigenmode.  This behavior
indicates a MnAs particle, whose hard axis points approximately in the
$r_1 r_2$-plane.  In fact, the features of the measured $\Delta f_i
(B)$ in Fig.~\ref{fig:2}(b) are reproduced by a simulation considering
a nearly symmetric half-ellipsoid with a hard-axis lying $\theta_K =
72^{\circ}$ from $\mathbf{\hat{n}}$ and $\phi_K = 6.3^{\circ}$ from
$\mathbf{\hat{r}_1}$.  These data are clear evidence that
crystallization of the liquid droplet can occasionally occur along a
direction far off of the NW growth axis.


In order to test the behavior of these NWs as scanning magnetic
sensors, we approach a typical one (NW2) to a current-carrying Au wire
patterned on a SiO$_2$ substrate, as described in Fig.~\ref{fig:4}(a).
Once in the vicinity of the wire constriction, the NW's two modes are
excited by the Biot-Savart field $\mathbf{B}_{\mathrm{AC}}$ resulting
from an oscillating drive current $I = I_1\sin(2\pi f_1t) +
I_2\sin(2\pi f_2t)$, where $I_1 = I_2 = 50 \mu\text{A}$.  Single
10~$\mu\text{m}$-long line scans are acquired by moving the NW across
the wire at the fixed tip-sample spacing $d_z = 250$~nm, while both
the resonant frequencies $f_i$ and displacement amplitudes $r_i$ are
tracked using two phase-locked loops.  The corresponding values of the
force driving each mode at resonance are then calculated as $F_i = r_i
k_i/Q_i$ (see Supplementary Information).

Using an approach similar to that used to calibrate MFM tips
\cite{lohau_quantitative_1999, schendel_method_2000,
  kebe_calibration_2004}, we model the force exerted by a well-known
magnetic field profile on the magnetic tip by using the so-called
point-probe approximation.  This approximation models the complex
magnetization distribution of the tip as an effective monopole moment
$q_0$ and a dipole moment $\mathbf{m}$ located at a distance $d$ from
the tip apex (the monopole contribution compensates for the
non-negligible spatial extent of the tip).  The magnetic force acting
on each mode is then given by $D_i = q_0 \mathbf{B}_{\mathrm{AC}}
\cdot \mathbf{\hat{r}}_i + \nabla(\mathbf{m}\cdot
\mathbf{B}_{\mathrm{AC}})\cdot \mathbf{\hat{r}}_i$.  Moreover, we also
consider the magnetic torque $\boldsymbol{\tau} = \mathbf{m} \times
\mathbf{B}_{\mathrm{AC}}$ generated at the tip, which results in a
torsion and/or bending of the NW depending on its orientation.
Although this contribution is negligible in conventional MFM, the NW
modes' short effective length ($l_e = 12.2\mu\text{m}$) and soft
spring constant ($k_1\approx k_2\approx 8.3\text{mN/m}$), make the
bending component of the torque responsible for an observable
displacement along $\mathbf{\hat{r}}_i$.  We then model the total
force driving each mode as $F_i = D_i + T_i$, where $T_i =
(l_e^{-1}\mathbf{\hat{n}} \times \boldsymbol{\tau}) \cdot
\mathbf{\hat{r}}_i$.  $d$ is set to $100$~nm from the tip (i.e.\
approximately at the base of the MnAs crystallite) for the best fits,
while $q_0$ and $\mathbf{m}$ are used as free parameters.  The precise
spatial dependence of the field $\mathbf{B}_{\mathrm{AC}}$ produced by
the current $I$ is calculated using the finite-element package
\textit{COMSOL} (see Supplementary Information).

We characterize the NW magnetic response at $B = -5\text{T}$ and
$B=0$.  In the high field case shown in Fig.~\ref{fig:4}(b), a fit of
the two driving forces is obtained with an effective dipole
$\mathbf{m}$ nearly along $\mathbf{B}$ with a magnitude close to the
dipole moment of a fully saturated tip: $|\mathbf{m}| = 0.73 M_sV$,
where $V = 1.62 \times 10^{-21}\;\text{m}^3$ is the volume of the tip
defined in the magnetometry simulation.  In general, the estimation of
$V$ from SEM is approximate due to the difficulty in determining the
precise three-dimensional geometry and in distinguishing regions of
non-magnetic material inside the tip or at its surface.  The fit
returns a small, but non-zero $q_0$ and shows a very good agreement
with the measured forces sensed by the two modes.  In
Fig.~\ref{fig:4}(c), the same line scan is performed in absence of
external field and, as expected, the observed response changes
radically, since the magnetization lies mostly in the easy plane and
is not completely saturated.  In the fit, the effective dipole
contribution is dominant and mostly orthogonal to the NW axis
$\mathbf{\hat{n}}$ with $|\mathbf{m}| = 0.53 M_sV$.  The direction of
$\mathbf{m}$ is closely related to the torque contribution making the
fit particularly sensitive to the value of $\phi_{\mathbf{m}}$.
Spurious electrostatic driving of the NW modes is negligible.


\begin{figure}[b!]
  \includegraphics[]{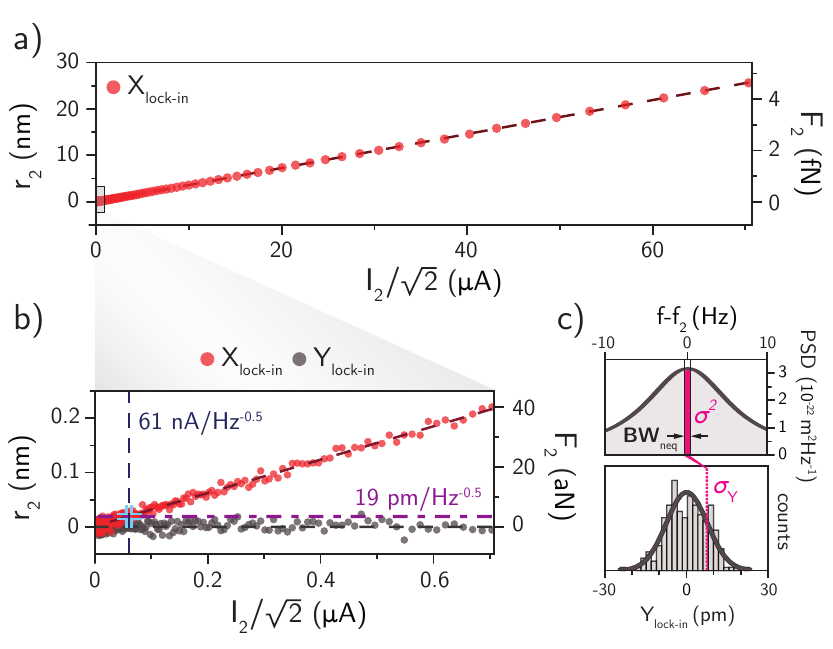}
  \caption{NW sensitivity to a resonant current drive at a distance of
    $250$~nm.  (a) Plot of the oscillation amplitude $r_2$ for each
    value of the current amplitude $I_2$, both quantities are root
    mean squared.  (b) Plot of in-phase response to the drive (X) and
    quadrature signal (Y) for a finer current sweep from $I_2 =
    1\;\mu$A to $I_2=1$ nA.  The lock-in demodulator low-pass filter
    noise equivalent bandwidth was set to $\text{BW}_{\text{neq}} =
    0.156$~Hz and each point was averaged for $2.5$ sec. Both signals
    are linearly fit (dark dashed lines).  The intersection (light
    blue cross) between the linearly fit signal $\bar{X}$ and the
    displacement sensitivity (dashed purple line), shows a current
    sensitivity of $61$~$\text{nA}/\sqrt{\text{Hz}}$ (dashed blue
    line).  (c) Noise analysis on the quadrature channel Y. On top, a
    close-up of a fit to the second mode's PSD in Fig \ref{fig:1}(c)
    and the noise power present within the measurement bandwidth
    $\text{BW}_{\text{neq}}$ (magenta box around $f_2$): $\sigma^2 =
    5.4\times10^{-23}\;\text{m}^2$ .  Below, the histogram of the
    noise measured on quadrature channel (Y) in (b), fitted by a
    Gaussian $\mathcal{N}(0,\sigma_{Y})$ with $\sigma_Y =
    7.5\;\text{pm} \sim \sqrt{\sigma^2}$, confirming the thermally
    limited nature of the sensitivity measurement.  }

\label{fig:5}
\end{figure}

The NWs' high force sensitivity combined with highly concentrated and
strongly magnetized dipole-like tips give them an exquisite
sensitivity to magnetic field gradients.  In order to quantify this
sensitivity, we restrict our attention to the second mode (red) of
NW2, positioning it at the point of maximal response over the wire at
$d_z = 250$ nm and $B=0$ T (\ie $y = 4.5\;\mu$m on Fig
\ref{fig:4}(c)).  The displacement signal $r_2$ is measured with a
lock-in amplifier while decreasing the driving current $I =
I_2\sin(2\pi f_2t)$.  The sweeps plotted in Fig.~\ref{fig:5}(a) and
(b) show the expected linear response as well as a wide dynamic range.
In Fig.~\ref{fig:5}(b), we focus on the low-current regime, showing
both the in-phase $X$ (signal+noise) and quadrature $Y$ (noise)
response.  By simple linear regression, we find a transduction factor
$\beta = 0.22$~$\text{nm}/\mu \text{A}$, where $\overline{X} = \beta
I_2$.  The noise in both $X$ and $Y$ is found to be Gaussian and fully
ascribable to the NW's thermal motion with variances $\sigma^2_X
\approx \sigma^2_Y \approx S_{r_2}(\omega_2)
\times\text{BW}_{\text{neq}}$, where $S_{r_2}(\omega)$ is a fit to the
second mode's thermal PSD shown in Fig.~\ref{fig:1}~(c), $\omega_2$ is
the resonant angular frequency of the second mode, and
$\text{BW}_{\text{neq}}$ is the lock-in's equivalent noise bandwidth.
As shown in Fig.~\ref{fig:5}~(c) for $Y$, the mode's thermal PSD is
assumed constant around its value at resonance $S_{r_2}(\omega_2) =
4k_BT\frac{Q_2}{k_2\omega_2}$, due to the narrow measurement
bandwidth.  Therefore, the NW's second mode has a thermally-limited
displacement sensitivity of $\sqrt{S_{r_2}(\omega_2)} =
19\;\text{pm}/\sqrt{\text{Hz}}$, equivalent to a force sensitivity of
$\sqrt{S_{r_2}(\omega_2)k_2/Q_2} = 3.5\;\text{aN}/\sqrt{\text{Hz}}$.
Given the measured current transduction factor $\beta$ at the working
tip-sample spacing $d_z = 250$~nm, we obtain a sensitivity to current
flowing through the wire of $61\;\text{nA}/\sqrt{\text{Hz}}$.


Such sensitivity to electrical current compares favorably to that of
other microscopies capable of imaging current through Biot-Savart
fields, including scanning Hall microscopy, magneto-optic microscopy,
scanning SQUID microscopy, microwave impedance microscopy, and
scanning nitrogen-vacancy
magnetometry~\cite{kirtley_fundamental_2010,chang_nanoscale_2017}.
Because of the dipole-like character of the MnAs tip, this
transduction of current into displacement is dominated by the effect
of the time-varying magnetic field gradient generated by the current:
$F_i \approx \nabla(\mathbf{m}\cdot \mathbf{B}_{\mathrm{AC}})\cdot
\mathbf{\hat{r}}_i = \mathbf{m} \cdot \nabla (\mathbf{B}_{\text{AC}}
\cdot \mathbf{\hat{r}}_i)$.  Although the torque resulting from the
time-varying magnetic field produces an effective force, $T_i$, as
seen in Figs.~\ref{fig:4}~(b) and (c), this term is typically
secondary.  Therefore, from \textit{COMSOL} simulations of the field
produced by current flowing through the wire, we find this current
sensitivity to correspond to a sensitivity to magnetic field gradient
of $11\;\text{mT}/(\text{m} \sqrt{\text{Hz}})$ at the position of the
tip's effective point probe, i.e.\ $d_z + d = 350$~nm above the
surface.  Having quantified the NW's response to magnetic field
gradients, we can calculate its sensitivity to other magnetic field
sources, including a magnetic moment (dipole field), a superconducting
vortex (monopole field), or an infinitely long and thin line of
current~\cite{kirtley_fundamental_2010}.  In particular, we expect a
moment sensitivity of 54~$\mu_B /\sqrt{\text{Hz}}$, a flux sensitivity
of 1.3~$\mu \Phi_0 /\sqrt{\text{Hz}}$, and line-current sensitivity of
$9\;\text{nA} / \sqrt{\text{Hz}}$.  These values show the capability
of magnet-tipped NWs as probes of weak magnetic field patterns and the
huge potential for improvement if tips sizes and tip-sample spacings
can be reduced (see Supplementary Information).

In addition to improved sensitivity, NW MFM provides other potential
advantages compared to conventional MFM.  First, scanning in the
pendulum geometry with the NW oscillating in the plane of the sample
has the characteristics of lateral MFM.  This technique, which is
realized with the torsional mode of a conventional cantilever,
distinguishes itself from the more commonly used tapping-mode MFM in
its ability to produce magnetic images devoid of spurious
topography-related contrast and in a demonstrated improvement in
lateral spatial resolution of up to
15\%~\cite{kaidatzis_torsional_2013}.  Second, the nanometer-scale
magnetic particle at the apex of the NW force sensor minimizes the
size of the MFM tip, allowing for optimal spatial resolution and
minimal perturbation of the investigated sample.

The prospect of increased sensitivity and resolution, combined with
few restrictions on operating temperature, make NW MFM ideally suited
to investigate nanometer-scale spin textures, skyrmions,
superconducting and magnetic vortices, as well as ensembles of
electronic or nuclear spins.  Non-invasive magnetic tips may also open
opportunities to study current flow in 2D materials and topological
insulators.  The ability of a NW sensor to map all in-plane spatial
force
derivatives~\cite{rossi_vectorial_2017,de_lepinay_universal_2017}
should provide fine detail of stray field profiles above magnetic and
current carrying samples, in turn providing detailed information on
the underlying phenomena.  Directional measurements of dissipation may
also prove useful for visualizing domain walls and other regions of
inhomogeneous magnetization.  As shown by Grutter et al., dissipation
contrast, which maps the energy transfer between the tip and the
sample, strongly depends on the sample's nanometer-scale magnetic
structure~\cite{grutter_magnetic_1997}.


\section*{Methods}

\textit{Interferometric detection}: The linearly polarized light
emitted by a laser diode with wavelength $\lambda = 1553$ nm is
directly coupled to a polarization maintaining optical fiber, sent
through the 5\% transmission arm of a 95:5 fiber-optic coupler,
collimated and focused on the NW by a pair of lenses. This confocal
reflection microscopy setup, analogous to the one described in
\cite{hogele_fiber-based_2008}, focuses light to a minimum beam waist
of $w_0 = 1.65\;\mu$m (see Supplementary Information).  The light
incident on the NW has a power of $25\;\mu$W and is polarized along
its long axis.  Light scattered back by the NW interferes with light
reflected by the fiber's cleaved end, resulting in a low-finesse
Fabry-Perot interferometer.  A fast photo-receiver monitors variations
in the intensity of reflected light, allowing for the sensitive
detection of NW motion.  The interferometric signal is proportional to
the projection of the NW's motion along the direction of the
interference pattern's gradient $\mathbf{\hat{m}}$.  The magnitude of
this gradient at the position of the NW determines the
interferometer's transduction factor.  By positioning the NW within
the optical waist or changing the wavelength of the laser, it is
possible to measure the motion projected along arbitrary directions.
For displacement measurements presented in this work, NWs have been
positioned on the optical axis $\mathbf{\hat{y}}$ in order to have
$\mathbf{\hat{m}} \parallel \mathbf{\hat{y}}$ (see
Supplementary Information).

\textit{Numerical simulations}: Micromagnetic simulations are carried
out with \textit{Mumax3}.  We set the saturation magnetization $\mu_0
M_s = 1.005$~T, the exchange stiffness constant $A = 10$~pJ/m, and the
magnetocrystalline anisotropy $K = -1.2 \times
10^6$~$\text{J}/\text{m}^3$ in correspondence with the values reported
for MnAs in the literature
\cite{de_blois_magnetic_1963,engel-herbert_micromagnetic_2006}.  We
model the geometry of each MnAs magnetic tip based on observations
made in a SEM. Space is discretized into cubic mesh elements, which
are $5$~nm on a side, which corresponds to the dipolar exchange length
of the material $l_{ex} = \sqrt{2A/(\mu_0M_s^2)}$. The validity of
this discretization is confirmed by comparing the results of a few
representative simulations with simulations using much smaller mesh
sizes.  \textit{Mumax3} determines the equilibrium magnetization
configuration for each external field value by numerically solving the
Landau-Lifshitz-Gilbert equation.  Since the microscopic processes in
a MnAs tip are expected to be much faster than the NW resonance
frequencies, the magnetization of the tip is assumed to be in its
equilibrium orientation throughout the cantilever oscillation.  The
calculation also yields the total magnetic energy $E_m$ corresponding
to each configuration.  In order to simulate $\Delta f_i$, we
numerically calculate the second derivatives of $E_m$ with respect to
$\theta_i$ found in (\ref{DCM_equation}).  At each field, we calculate
$E_m$ at the equilibrium angles $\theta_i = 0$ and at small deviations
from equilibrium $\theta_i = \pm \delta \theta_i$.  For small $\delta
\theta_i$, the second derivative can be approximated by a finite
difference: $\left.  \frac{\partial^2 E_m}{\partial \theta_i^2} \right
|_{\theta_i=0} \approx \frac{E_m(\delta \theta_i) - 2 E_m(0) +
  E_m(-\delta \theta_i)}{(\delta \theta_i)^2}$. By setting $f_{0_i}$,
$k_i$, and $l_e$ to their measured values, we then arrive at the
$\Delta f_i$ corresponding to each magnetization configuration in the
numerically calculated field dependence.

\section*{Supplementary Information}

Supplementary information is available for download at
\href{http://poggiolab.unibas.ch/full/dropbox/RossiSupp.pdf}{RossiSupp.pdf}.
This file includes sections discussing: the optical setup for NW
motion detection; displacement calibration and force estimation;
micromagnetic simulations of NW magnetometry; simulated
$B_{\text{AC}}$ profile; sensitivity to different types of magnetic
field sources.

Movies of the simulated magnetization reversal for the 
presented MnAs tips are available at
\href{http://poggiolab.unibas.ch/full/dropbox/NW1.avi}{NW1.avi},
\href{http://poggiolab.unibas.ch/full/dropbox/NW2.avi}{NW2.avi},
\href{http://poggiolab.unibas.ch/full/dropbox/NW3.avi}{NW3.avi}.
\section*{Acknowledgements}

We thank Sascha Martin and his team in the machine
shop of the Physics Department at the University of Basel for help
building the measurement system.  We thank Lorenzo Ceccarelli for help
with magnetic field simulations and Prof. Ilaria Zardo for suggesting
the collabroation.  We acknowledge the support of the Kanton Aargau,
the ERC through Starting Grant NWScan (Grant No. 334767), the SNF
under Grant No. 200020-178863, the Swiss Nanoscience Institute, the
NCCR Quantum Science and Technology (QSIT), and the DFG via project
GR1640/5-2 in SPP 153.

%




%


\end{document}